# An Electrocardiogram Monitoring Device Based on STM32

Wenqi Guan [1*]

[1] School of Electronic Science and Engineering, Nanjing University, Nanjing, China, 210023

*Corresponding author: 221180006@smail.nju.edu.cn

**Abstract:** Cardiovascular diseases remain the leading cause of morbidity and mortality, particularly in aging populations, with rising rates of heart-related incidents in middle-aged and even younger individuals. Currently available electrocardiogram (ECG) monitoring technologies, such as bedside monitors are bulky and cumbersome, limiting their potential accessibility for continuous, real-time monitoring. Therefore, we propose a portable digital ECG monitoring system that can accurately measure heart rate, analyze the ECG in real-time, and wirelessly send data to the cloud platform. The device comprises a custom-designed circuit to capture ECG signals, with analog-to-digital conversion and heart rate measurement performed by an STM32F429 microcontroller. In addition, real-time monitoring is achieved through the L610-4G module, which transmits data to Tencent Cloud, enabling users to access their ECG data remotely while being notified by instant alerts for abnormal heart rates. Moreover, healthcare professionals can view historic ECG data for further clinical analysis of patient progression and make evidence-based decisions. The proposed system features portability and accuracy, enabling the user to access extended ECG data easily. It provides the ability to minimize redundant clinical assessments, thus offering considerable advantages to patients and healthcare practitioners.

**1. Introduction**

Cardiovascular health monitoring has become an essential part of modern healthcare, with the demand for real-time, accurate, and easily accessible heart data growing rapidly [1]. Traditional ECG devices provide basic heart activity monitoring but often lack advanced features such as seamless cloud integration, real-time feedback, and automated alerts. To address these limitations, we designed an ECG monitoring system based on the STM32 microcontroller that captures, processes, and analyzes heart signals with high precision, while offering advanced connectivity and cloud-based functionality.

Our system connects ECG leads to the user's left hand and right leg, capturing signals that pass through several amplification and filtering stages, including high/low-pass filters, a 50Hz notch filter, and a right-leg drive circuit. The amplified ECG signal (~2Vpp with a DC bias) is processed by the STM32 microcontroller's ADC using a DMA ping-pong structure for continuous data acquisition. The microcontroller analyzes the ECG signal frequency and transmits the results via a serial interface to the L610-4G module, which uploads the data to Tencent Cloud. Users are then provided with real-time notifications and alerts based on the analysis, enabling timely responses to any abnormalities.

In addition to its core functionality, the device supports extended features such as cloud-based ECG waveform rendering [2], voice alerts for abnormal heart rates, and digital filtering to reduce noise and enhance signal clarity. These features make our system a powerful tool for both patients and healthcare professionals, promoting real-time monitoring and remote diagnostics, which are crucial for proactive heart health management.

The contributions of this design include custom-designed analog signal acquisition and processing circuits, digital filtering and ADC functions based on the STM32 microcontroller, along with the integration of a communication module to enable ECG waveform plotting and following QRS complex detection.

## 2. Construction of Specimens

The overall system flow is illustrated in Figure 1. The primary task of the hardware circuitry is to amplify and filter the weak ECG signals for microcontroller acquisition. The signal needs to be amplified by approximately 1500 times, and the frequency is restricted within the range of 0.05Hz to 70Hz, with a 50Hz notch filter applied to suppress power line interference [3]. The signal undergoes voltage lifting, ensuring the output voltage remains within 0 to 3.3V. Additionally, a right-leg drive feedback circuit is employed to further reduce interference and measurement errors caused by the human body.

For the software aspect, once the microcontroller's ADC captures the data, digital filtering is applied to further eliminate 50Hz power line and electromyogram (EMG) interference. The system then plots the ECG waveform on an OLED screen and calculates the heart rate using edge-triggered detection. The heart rate data is transmitted via a serial interface to the L610 module and uploaded to Tencent Cloud. By configuring Tencent Cloud's API, the heart rate can be displayed on the Tencent WeChat public account and mini-program. If the heart rate exceeds preset thresholds, the L610 module triggers a voice alert and sends real-time notifications to both the user and emergency contacts, including the abnormal heart rate, an alert message, and the user's current location.

In the subsequent phase, the ECG data collected by the microcontroller will be uploaded to Tencent Cloud via the L610 module. By calling APIs in Tencent Cloud through Python, the stored ECG data can be retrieved and visualized using the Matplotlib library. This enables real-time remote monitoring of the patient's heart activity from any location.

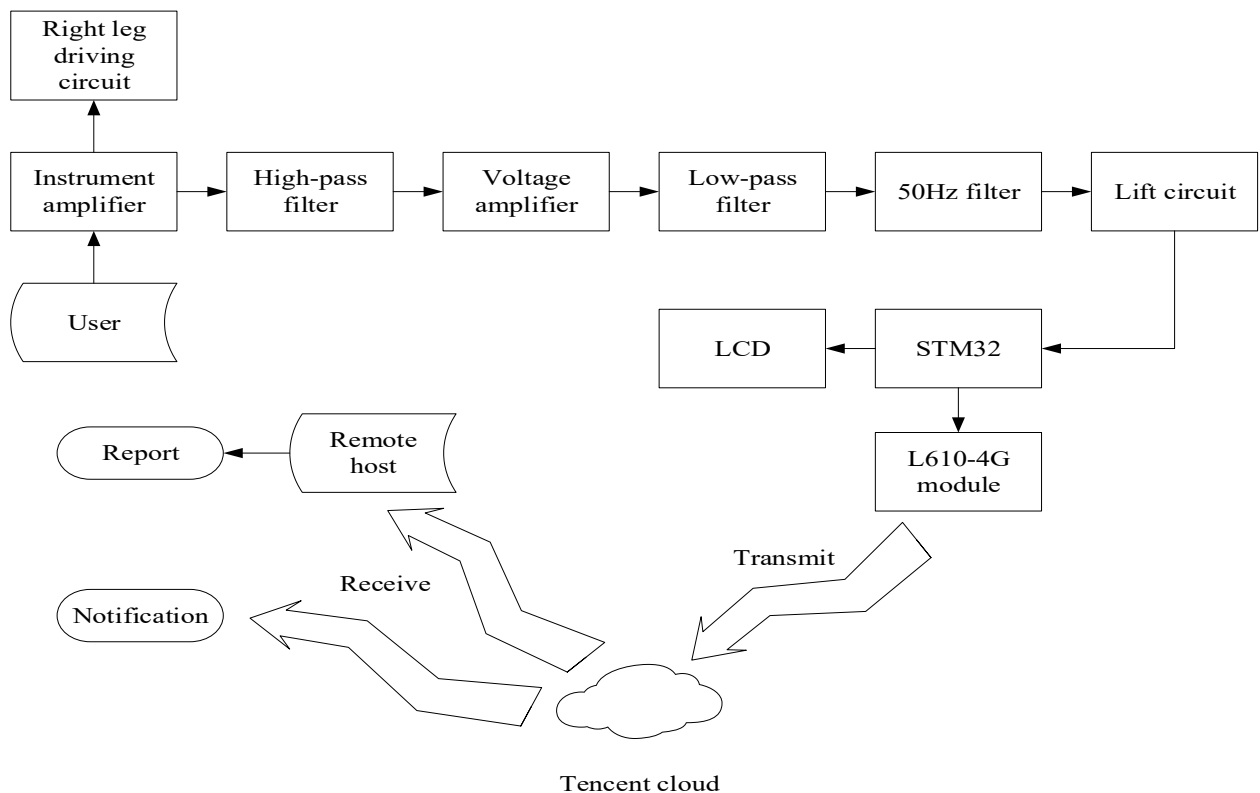

Figure 1. ECG structure

### 2.1 Hardware circuitry

The analog signal acquisition part consists of seven basic elements, which are:

(1) The Instrument Amplifier is responsible for the initial amplification of weak input signals, ensuring high input impedance to avoid signal distortion. It amplifies the differential voltage between two input signals while suppressing any common-mode noise. This design makes it particularly suitable for bio-signal measurements, such as ECG, where accurate and low-noise signal amplification is critical.

In this part, a precise amplifier (Vos < $1mV$) OP07CP (shown in Figure 2., left panel) is selected. It provides exceptional low-noise performance, with an input voltage noise density of just 10.5 nV/$\sqrt{Hz}$, which is crucial for accurately amplifying weak signals such as the ECG without introducing noise that could obscure important physiological data. The device's low offset voltage and excellent long-term stability further ensure that the amplified signals remain accurate over time, even in precision applications. Additionally, the OP07CP is built using bipolar transistor technology, which enhances the common-mode rejection ratio (CMRR) and makes it ideal for applications that demand high precision in noisy environments.

The OP07CP also operates over a wide range of supply voltages ($\pm 3V$ to $\pm 22V$), providing flexibility in system design and allowing it to be easily integrated with the other components in the system. Moreover, its low bias current and high input impedance ensure that it does not interfere with the low-power signals typical of biological monitoring. Moreover, its internal compensation for frequency and low thermal drift makes the OP07CP highly reliable for continuous monitoring applications where long-term stability is required [4]. The device's relatively wide gain bandwidth (600 kHz) ensures sufficient amplification speed for ECG signal processing, while the wide input voltage range further contributes to the system's ability to handle variations in signal amplitude effectively. These factors collectively make the OP07CP a powerful and reliable choice for the instrumentation amplifier in this monitoring system.

(2) The Voltage Amplifier provides further gain to ensure the signal is strong enough for processing. It is designed to preserve the integrity of the signal while boosting its amplitude for downstream filtering and conversion stages.

In this and the following sections, the TL084 operational amplifier (shown in Figure 2., right panel) plays an essential role, attributing to its high-speed JFET input stage, which makes it particularly suitable for various stages of signal processing in the ECG system. Its high input impedance and low input bias current (typically 65 dB CMRR) make it ideal for amplifying small, high-impedance signals generated in biomedical applications [5]. Moreover, it has a low input offset current (typically 10 pA), which helps reduce errors and maintains high precision when working with low-level signals behind the Instrument Amplifier.

The slew rate of the TL084 is one of its standout features, measured at 16 V/μs, which enables fast response to quick signal changes without distortion or delay. A high slew rate also ensures excellent performance in the filtering stages, allowing precise rejection of unwanted noise while maintaining signal fidelity.

For versatile and reliable functions, the TL084 provides a wide differential input voltage range (up to $\pm 15V$), along with output short-circuit protection, it also boasts a gain bandwidth product of 4 MHz, providing high-frequency stability that is crucial for maintaining signal integrity across a broad frequency spectrum. Additionally, with a power supply rejection ratio (PSRR) of 75 dB and total harmonic distortion of 0.003% [6], the TL084 maintains performance consistency across a wide range of operating conditions. These attributes are particularly beneficial for the filtering and amplification circuits in the system, where clean and accurate signal reproduction is vital.

(3) The High-Pass Filter The high-pass filter removes low-frequency noise, such as baseline wander and motion artifacts. By setting a cutoff frequency of around 0.05 Hz, it effectively filters out any unwanted DC components from the ECG signal while preserving the relevant physiological frequencies.

(4) The Low-Pass Filter is to eliminate high-frequency noise, such as muscle artifacts and external electronic interference, the low-pass filter sets an upper cutoff at 70 Hz. This allows the system to focus on the frequency range of the heart's electrical activity, which typically falls below this threshold.

(5) The 50Hz Notch Filter is a specialized filter designed to suppress power-line interference, which is common in many medical and industrial settings. The notch filter specifically attenuates signals at 50 Hz to mitigate the influence of mains power noise.

(6) The Voltage Lifting Circuit is used to properly interface the amplified ECG signal with the ADC in the microcontroller, this circuit provides a suitable DC bias and gain adjustment, ensuring the signal fits within the 0–3.3V input range of the microcontroller's ADC.

(7) The Right Leg Drive Circuit enhances the common-mode rejection of the ECG system by feeding back an inverted common-mode signal to the patient's right leg. This reduces interference, such as power-line noise, and improves the signal-to-noise ratio by minimizing common-mode signals from the body [7].

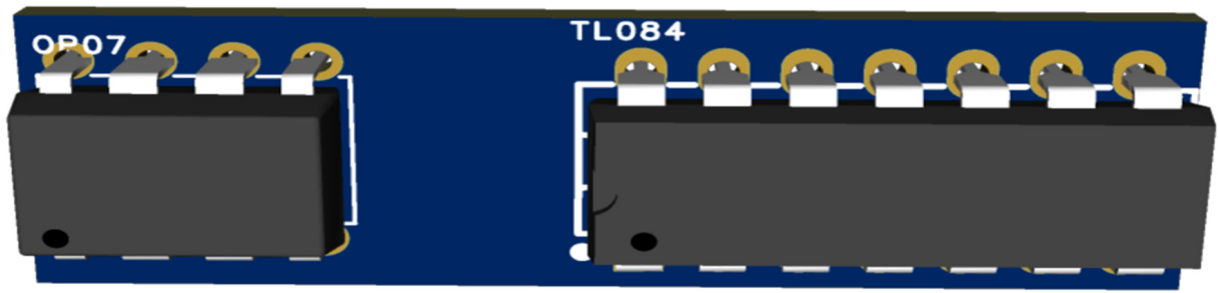

Figure 2. Operational amplifier

## 2.2 Software

The STM32F429 microcontroller, built on the ARM Cortex-M4 core, is an excellent choice for a range of applications in the ECG system due to its balance of performance, power efficiency, and integration features [8]. Running at 180 MHz, it delivers high processing capability, executing from Flash memory with zero wait states thanks to the integrated ART Accelerator, reaching a peak performance of 225 DMIPS and 608 CoreMark. This makes it especially effective for handling complex computations. Additionally, thanks to STMicroelectronics' advanced 90 nm process, achieving a current consumption as low as 260 μA/MHz when running at full speed (180 MHz), which is crucial for portable medical devices where battery life is a concern. The low-power Stop mode consumes only 120 μA, allowing the system to maintain low energy consumption during idle periods.

In my design, ADC sampling and signal analysis are key, the STM32F429's three 12-bit ADCs, reaching up to 2.4 MSPS or 7.2 MSPS in interleaved mode, provide high-speed and accurate signal conversion of the ECG data. This enables the real-time acquisition and processing of signals, making it suitable for continuous health monitoring. Furthermore, the 12-bit DACs are available for outputting signals if needed for feedback or interfacing with external devices.

For the curve plotting task, its LCD-TFT controller with Chrom-ART Accelerator allows users to efficiently render the ECG waveform on an LCD display [9]. The dual-layer support and image processing functionalities like image blending and format conversion reduce the load on the core, ensuring smooth display performance without compromising the microcontroller's real-time signal processing tasks. Besides, numerous communication interfaces, including I²S and USARTs (running at up to 11.25 Mbit/s), allowing flexible interfacing with external devices (e.g. L610-4G).

## 3. Experiments

### 3.1 The establishment of an analog signal acquisition circuit

The signal is captured from the left wrist, where the amplitude is very small (only a few microvolts). Given the presence of common-mode interference, a reference signal is also taken from the right leg to suppress this noise. The captured differential signal, with an amplitude of around 1 mV, is first processed by an instrument amplifier, which amplifies the differential signal while effectively rejecting the common-mode interference due to its high common-mode rejection ratio (CMRR).

For the circuit design, surface-mount resistors and capacitors are utilized to allow for fine adjustments during tuning. The OP07 is chosen for the instrument amplifier because it offers low noise and high precision, which are critical for the weak input signal, ensuring maximum CMRR at the input stage. In subsequent stages, the TL084 is selected for its high slew rate and cost-efficiency, integrating four operational amplifiers in a single package, which not only simplifies the PCB design but also reduces cost. The DIP-14 package allows for easy replacement in case of malfunction during field use.

Each module in the circuit is connected via $1 \times 3$ pin headers, where the second and third pins are shorted, and the third serves as a test point. The first pin connects to the next stage of the circuit, and a jumper can short the first and second pins after successful debugging. This modular design allows for isolated testing of each section, simplifying troubleshooting and ensuring the integrity of individual components before integration. Before the actual PCB design, theoretical calculations are carried out to estimate the expected parameters (Variable names are shown in Figure 3., left panel), followed by simulation in Multisim. Specific parameter calculations are shown below (Only basic static parameters are calculated here):

$$A_{Instrument} = A_1 \times A_2 = \left(1 + \frac{R_3+R_4}{R_1+R_2}\right) \times \frac{R_7}{R_5} = 22 \qquad (1)$$

$$f_{ch} = \frac{1}{2\pi C_2 A_{11}} = 0.072 Hz \qquad (2)$$

$$f_{cl} = \frac{1}{2\pi C_3 R_{15}} = 70.73 H_2 \qquad (3)$$

$$f_0 = \sqrt{f_{4p} \times f_{4p}} = \sqrt{\frac{1}{2\pi R_{31} C_5} \times \frac{1}{2\pi R_{27} C_7}} = 49.79 Hz \qquad (4)$$

Once the simulation confirms the expected behavior, the circuit is designed in an EDA tool, where the schematic is translated into a PCB layout for component placement and signal routing. After fabrication, the PCB undergoes soldering for component assembly and further testing.

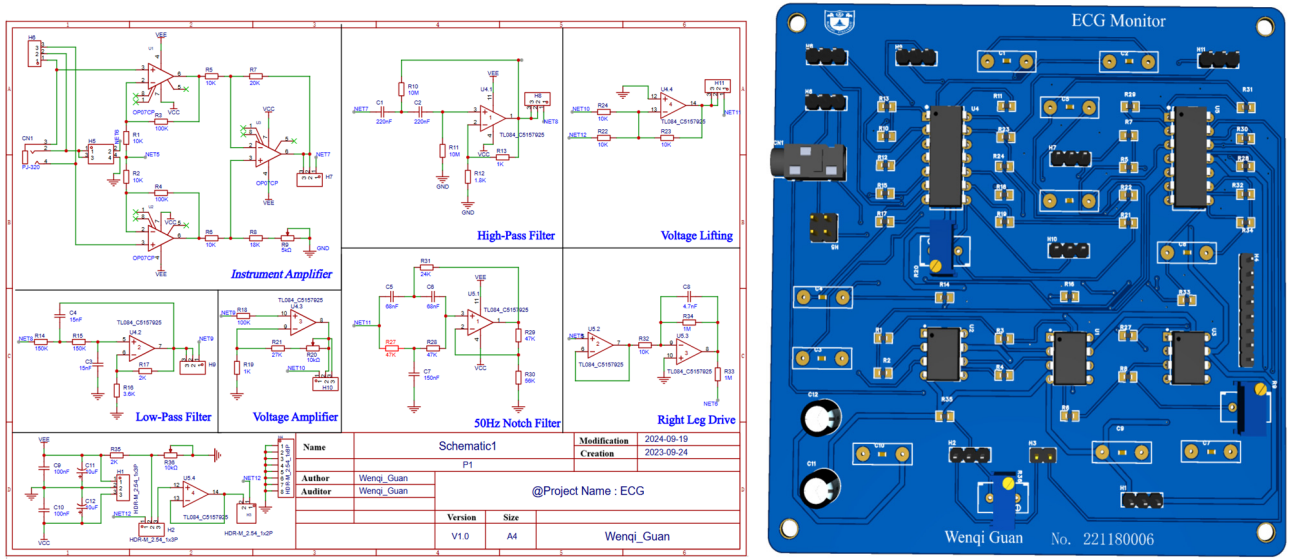

Figure 3. Schematic and PCB

### 3.2 Heart rate acquisition and STM32 system setup

PCB customization, component soldering, and wiring are shown in Figure 4., where the left side is a comparison image before and after PCB soldering and wiring, and the right side shows the pin diagrams of the OP07 and TL084.

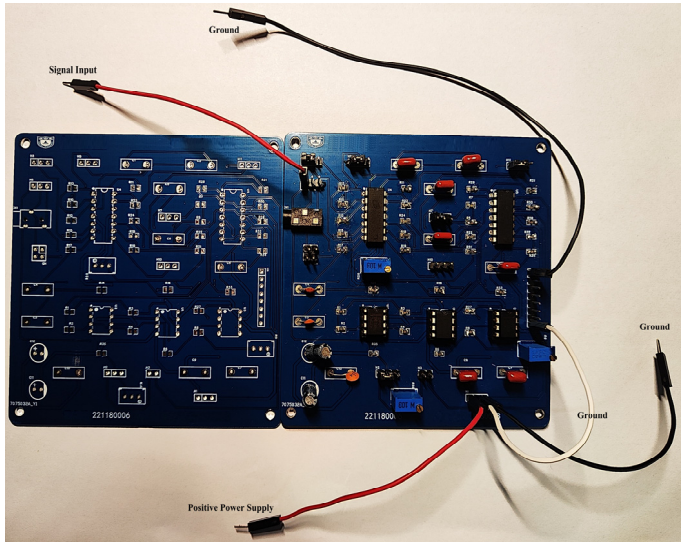
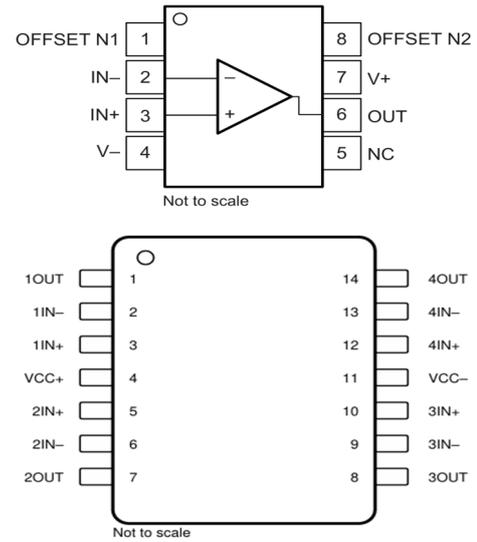

Figure 4. PCB wiring and chip pin diagram

Figure 5. the left panel presents the framework for signal acquisition and ADC in STM32. The ECG signal is captured using a single-channel setup, where a timer-triggered ADC is employed for data acquisition, and DMA is used to transfer the data to improve MCU efficiency. An efficient ping-pong structure for ADC acquisition and DMA data transfer (shown in Figure 5. right panel), ensuring continuous availability of an address for DMA operations [10]. While one buffer stores data for tasks such as algorithm processing, waveform rendering, and feature analysis, the other buffer is used for real-time data transfer. This ping-pong approach can be considered a form of pipeline technique, where the incoming data stream is divided evenly between two data buffers through a selection unit.

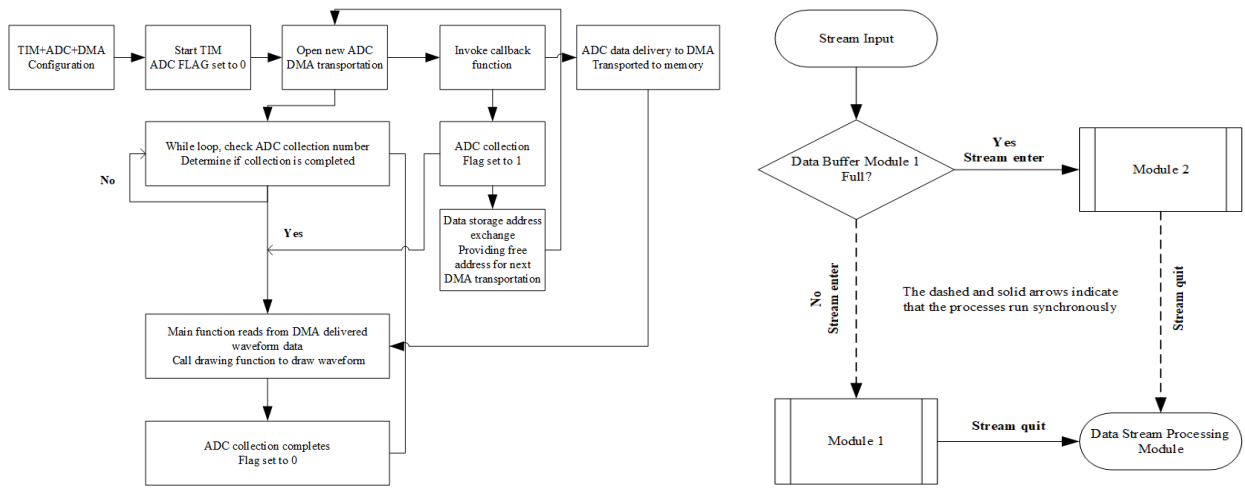

Figure 5. ADC framework and Ping-Pong structure

Additionally, we aim to eliminate 50Hz power line interference and EMG noise through digital filtering. The 50Hz filtering can be achieved using FFT by adjusting the amplitude of the corresponding harmonic components in the ECG signal. The basic concept for waveform display involves converting the digitized signals collected by the ADC into coordinate values for the OLED screen, storing them in an array, and connecting the points to form a continuous line. Before drawing new lines, previously drawn lines are erased by overwriting them in white, ensuring a smooth display transition. For heart rate calculation, the method is based on detecting the trigger points, such as the first rising edge, the falling edge, and the subsequent second rising edge of the waveform. To detect trigger points, the algorithm begins by scanning the signal from a specified position, searching for three consecutive monotonically increasing or non-decreasing data points near the trigger voltage. In most cases, this indicates the detection of a rising edge, and the middle point of these data is identified as the rising edge trigger point.

Generally, the time interval between two rising edges corresponds to one cycle of the waveform, and the frequency is determined from the time difference between these two points. For ECG waveforms, the frequency can be converted into heart rate accordingly.

## 4. Results

By cascading the analog signal acquisition circuit, STM32F429, and the L610-4G module (shown in Figure 7. left panel), we successfully implemented heart rate acquisition, analog-to-digital (AD) conversion, and real-time ECG waveform rendering. Table 1. presents the actual performance of the signal acquisition circuit, while Figure 6. compares the signal quality before and after incorporating the right-leg drive circuit (left side) and demonstrates the STM32's ability to measure and display the output from a 2Hz ECG signal generator in real-time (right side).

Table 1. Measuring Result

| Metrics | Ideal values | Actual values | Relevant formulas |
|---|---|---|---|
| Differential gain (Full) | >1500 | 1650 | $A_d = \dfrac{U_0}{U_{id}}$ |
| Input impedance | $\geq 5M\Omega$ | $13.2M\Omega$ | $Z_{in} = \dfrac{V_{in}}{I_{in}}$ |
| Short-circuit noise | $<0.1mV$ | $15.0\mu V$ | $e_n = \dfrac{U_{omax}}{A_d}$ |
| Bandwidth range | $0.2\sim 70Hz$ | $0.18\sim 70.2Hz$ | $BW = f_h - f_L$ |
| CMRR | $\geq 60dB$ | $93.16dB$ | $CMRR = 20 \cdot \lg \dfrac{A_d}{A_C}$ |
| 50Hz attenuation | $\leq -5dB$ | $-12.6dB$ | $\alpha = 20 \cdot \lg (\dfrac{A_{50Hz}}{A_{20Hz}})$ |

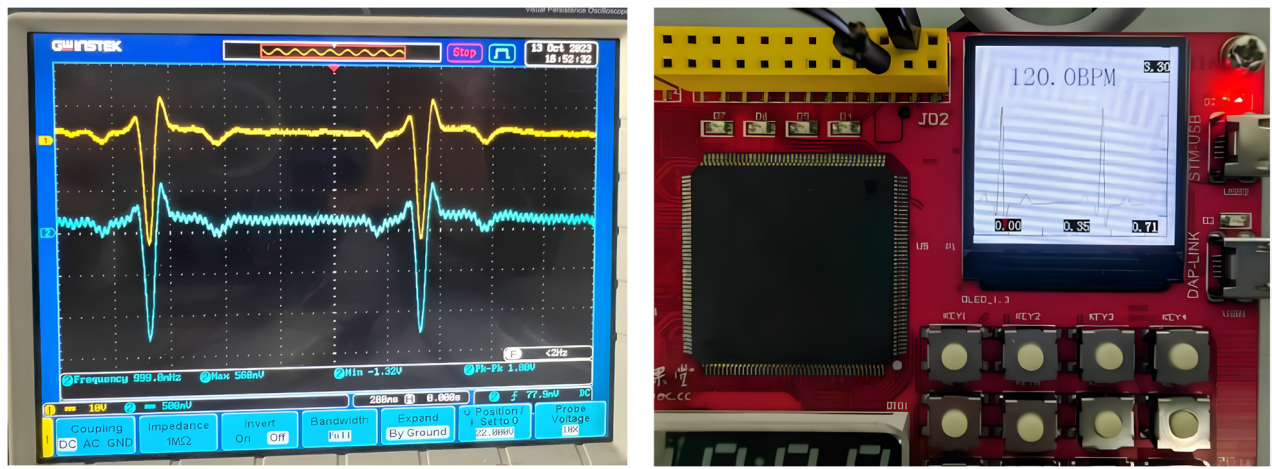

Figure 6. Comparison and Real test on STM32F429

By utilizing serial debugging tools (such as SSCOM), data communication between the L610-4G module and the cloud platform can be achieved. The backend can retrieve real-time ECG data stored as array variables on the cloud via API calls, formatted as JSON files. These data can then be processed to plot real-time ECG waveforms using Python. In the right panel of Figure 7, the real-time ECG waveforms are shown for a test subject, with the ECG monitor connected to both wrists and between the left wrist and right leg, respectively.

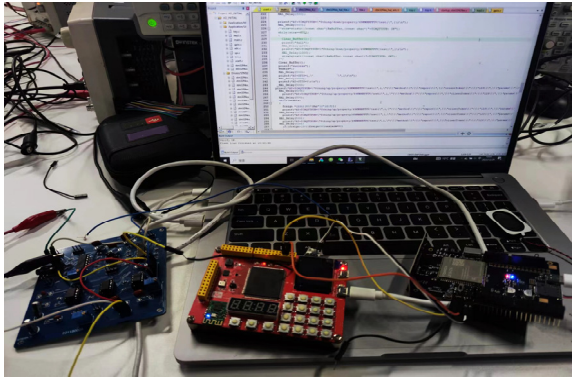

Figure 7. System overview and ECG plotting

## 5. Conclusions

This project developed an ECG monitoring system, leveraging advanced signal acquisition and processing techniques to ensure accurate heart rate measurement and real-time ECG waveform display. The system incorporated an analog circuit and a microcontroller for ADC conversion as well as data transmission via a 4G module to cloud storage. Through modular design and careful component selection (the OP07 for its high CMRR and the TL084 for its high slew rate), we ensured both reliability and cost-efficiency. The experimental results demonstrate that the system can achieve accurate signal acquisition and processing in real-time, offering robust performance in medical monitoring applications. This design showcases the feasibility of integrating analog and digital systems for portable healthcare devices, highlighting its value for continuous cardiac monitoring.

## References

[1] Merdjanovska E, Rashkovska A. Comprehensive survey of computational ECG analysis: Databases, methods and applications[J]. Expert Systems with Applications, 2022, 203: 117206.

[2] Mian Qaisar S, Subasi A. Cloud-based ECG monitoring using event-driven ECG acquisition and machine learning techniques[J]. Physical and Engineering Sciences in Medicine, 2020, 43(2): 623-634.

[3] Anbalagan T, Nath M K, Vijayalakshmi D, et al. Analysis of various techniques for ECG signal in healthcare, past, present, and future[J]. Biomedical Engineering Advances, 2023, 6: 100089.

[4] Güvenç H. Wireless ECG device with Arduino[C]//2020 Medical Technologies Congress (TIPTEKNO). IEEE, 2020: 1-4.

[5] NAIDU N, PANDEY P. Hardware for impedance cardiography [J]. M Tech Dissertation, 2005.

[6] TRIWIYANTO E Y, LAMIDI M R M R. Recent technology and challenge in ECG data acquisition design: A review; proceedings of the 2021 International Seminar on Application for Technology of Information and Communication (iSemantic), F, 2021 [C]. IEEE.

[7] Yang W. A new type of right-leg-drive circuit ECG amplifier using new operational amplifier[C]//Journal of Physics: Conference Series. IOP Publishing, 2021, 1846(1): 012034.

[8] MEWADA H K, DEEPANRAJ B. Low-Power Embedded ECG Acquisition System for Real-Time Monitoring and Analysis; proceedings of the 2024 IEEE World AI IoT Congress (AIIoT), F, 2024 [C]. IEEE.

[9] ZHAO Y, LIU Y, ZHOU H, et al. Intelligent Gateway Based Human Cardiopulmonary Health Monitoring System [J]. Journal of Sensors, 2023, 2023(1): 3534224.

[10] ZHOU W, YANG S. Optimization design of high-speed data acquisition system based on DMA double cache mechanism [J]. Microelectronics Journal, 2022, 129: 105577.